\documentclass[aps,prl,twocolumn,floatfix,amsmath,ams,showpacs]{revtex4-1}
\usepackage{graphicx}
\usepackage{amsmath,amssymb}
\usepackage{amsfonts}
\usepackage{mathrsfs}
\usepackage{epsfig}
\usepackage{dcolumn}
\usepackage{enumerate}
\usepackage{amsthm}

\begin{document}

\title{On the number of real eigenvalues of products of random matrices \\ and an application to quantum entanglement}
\author{Arul Lakshminarayan\footnote{e-mail: arul@physics.iitm.ac.in}}
\affiliation{Department of Physics, Indian Institute of Technology Madras, Chennai, 600036, India}

\begin{abstract}
The probability that there are $k$ real eigenvalues for an $n$ dimensional real random matrix is known. Here we study this for the case of products of independent random matrices. Relating the problem of the probability that the product of two real 2 dimensional random matrices has real eigenvalues to an issue of optimal quantum entanglement, this is fully analytically solved. It is shown that in $\pi/4$ fraction of such products the eigenvalues are real. Being greater than the corresponding known probability ($1/\sqrt{2}$) for a single matrix, it is shown numerically that the probability that {\it all} eigenvalues of a product of random matrices are real tends to unity as the number of matrices in the product increases indefinitely. Some other numerical explorations, including the expected number of real eigenvalues is also presented, where an exponential approach of the expected number to the dimension of the matrix seems to hold.

\end{abstract}
\pacs{02.50.-r, 03.67.-a}

\maketitle

\newmuskip\pFqskip
\pFqskip=6mu
\mathchardef\pFcomma=\mathcode`, 

\newcommand*\pFq[5]{%
  \begingroup
  \begingroup\lccode`~=`,
    \lowercase{\endgroup\def~}{\pFcomma\mkern\pFqskip}%
  \mathcode`,=\string"8000
  {}_{#1}F_{#2}\biggl[\genfrac..{0pt}{}{#3}{#4};#5\biggr]%
  \endgroup
}
\newcommand{\newc}{\newcommand}
\newc{\beq}{\begin{equation}}
\newc{\eeq}{\end{equation}}
\newc{\kt}{\rangle}
\newc{\br}{\langle}
\newc{\beqa}{\begin{eqnarray}}
\newc{\eeqa}{\end{eqnarray}}
\newc{\pr}{\prime}
\newc{\longra}{\longrightarrow}
\newc{\ot}{\otimes}
\newc{\rarrow}{\rightarrow}
\newc{\h}{\hat}
\newc{\bom}{\boldmath}
\newc{\btd}{\bigtriangledown}
\newc{\al}{\alpha}
\newc{\be}{\beta}
\newc{\ta}{\theta}
\newc{\ld}{\lambda}
\newc{\sg}{\sigma}
\newc{\p}{\psi}
\newc{\eps}{\epsilon}
\newc{\om}{\omega}
\newc{\mb}{\mbox}
\newc{\tm}{\times}
\newc{\hu}{\hat{u}}
\newc{\hv}{\hat{v}}
\newc{\hk}{\hat{K}}
\newc{\ra}{\rightarrow}
\newc{\non}{\nonumber}
\newc{\ul}{\underline}
\newc{\hs}{\hspace}
\newc{\longla}{\longleftarrow}
\newc{\ts}{\textstyle}
\newc{\f}{\frac}
\newc{\df}{\dfrac}
\newc{\ovl}{\overline}
\newc{\bc}{\begin{center}}
\newc{\ec}{\end{center}}
\newc{\dg}{\dagger}
\newc{\prh}{\mbox{PR}_H}
\newc{\prq}{\mbox{PR}_q}
\newc{\tr}{\mbox{tr}}
\newc{\pd}{\partial}
\newc{\qv}{\vec{q}}
\newc{\pv}{\vec{p}}
\newc{\dqv}{\delta\vec{q}}
\newc{\dpv}{\delta\vec{p}}
\newc{\mbq}{\mathbf{q}}
\newc{\mbqp}{\mathbf{q'}}
\newc{\mbpp}{\mathbf{p'}}
\newc{\mbp}{\mathbf{p}}
\newc{\mbn}{\mathbf{\nabla}}
\newc{\dmbq}{\delta \mbq}
\newc{\dmbp}{\delta \mbp}
\newc{\T}{\mathsf{T}}
\newc{\J}{\mathsf{J}}
\newc{\sfL}{\mathsf{L}}
\newc{\C}{\mathsf{C}}
\newc{\B}{\mathsf{M}}
\newc{\V}{\mathsf{V}}
\newc{\ovmg}{\overline{MG}}



While the study of the spectra of random matrices has been extensive and applications have been too numerous and varied to 
state briefly \cite{Forrester2003}, that of products of random matrices is relatively fewer, even though it is well-motivated \cite{ProductsRMTSP,ProductsRMTSE,ProductsRMTAM}. For example, products of random matrices could describe Jacobian matrices of chaotic systems and the rate of exponential increase of the largest eigenvalue gives the Lyapunov exponent. A similar consideration arises for problems involving disordered systems, with the matrices being transfer operators and the Lyapunov exponent being localization lengths. Recent studies of the spectral properties of products of random matrices include \cite{Burda2012,RogaZycz2011}.

The Ginibre ensemble of random matrices is the simplest to construct, as these are $n \times n$ matrices with all the $n^2$ entries being i.i.d. random variables drawn from a normal distribution such as $N(0,1)$ with zero mean and unit variance \cite{Ginibre1965}. If the entries are complex, the real and imaginary parts are independent random variables. For the purposes of this work, attention is restricted to the real ensemble. It is known that the eigenvalues of such matrices can have a significant fraction of eigenvalues that are themselves real. Explicit expressions for $p_{n,k}$, the probability that $k$ eigenvalues are real for a random $n\times n$ real matrix, have been found. Although these are not simple, there are elegant formulae for $E_n$, the expected number of real eigenvalues as well as the probability that there are exactly $n$ real eigenvalues \cite{Edelman1994,Edelman1997,Kanzieper2005}. For example it is known that: $\lim_{n \rightarrow \infty}E_n/\sqrt{n} =\sqrt{2/\pi}$ and $p_{n,n}=2^{-n(n-1)/4}$ \cite{Edelman1994}.

It is interesting therefore to study of the number of real eigenvalues of products of random matrices. If there are $K$ matrices in the product, let the probability that it has $k$ real eigenvalues be denoted by $p^{(K)}_{n,k}$. It is shown below that $p^{(2)}_{2,2}=\pi/4$, and is therefore larger than the probability that a  $2$ dimensional random matrix has real eigenvalues, which is $p_{2,2}=1/\sqrt{2}$. Numerical results indicate that $p^{(K)}_{2,2}$ monotonically increases to $1$ as $K$ increases to $\infty$, thus the probability that there are real eigenvalues increases with the number of matrices in the product. Numerical results also indicate identical conclusions for matrices of dimensions larger than $2$, namely that $p^{(K)}_{n,n}$ is a monotonically increasing function of $K$ and seems to tend to unity. The distribution of the matrix elements for $K>1$ are not naturally not independent, but that the correlations lead to this is a somewhat surprising result. Readers not interested in quantum entanglement may go directly to the paragraph following Eq.~(\ref{prod2}).

One direct application of the result for $p^{(2)}_{2,2}$ to a problem in quantum entanglement \cite{Horodecki2009} is to find the fraction of real ``optimal" states \cite{Wootters1998,Uhlmann2010,Shuddhodhan2011} of rank 2. 
 A set of pure states of two qubits $\{|\phi_i\kt, \, i=1, \ldots, k \}$ are C-optimal if for any $\{p_i, \; i=1, \ldots,k,\; \sum_i p_i =1, \; p_i\ge 0\}$ one has:
\beq
\label{coptimal}
C\left(\rho=\sum_{i=1}^k p_i |\phi_i \kt \br \phi_i|\right) = \sum_{i=1}^k p_i C\left(|\phi_i \kt \br \phi_i|\right),
\eeq
$C(\cdot)$ being the concurrence function \cite{HillWootters1997,Wootters1998}, a measure of entanglement between the two qubits. In general the R.H.S. is larger than the L.H.S., the concurrence being a convex function, and in this sense the set of states leads to optimally entangled mixtures if the equality is satisfied.

Restricting oneself to the set of states that are {\it real} in the standard basis, it was shown in \cite{Shuddhodhan2011} that when $k=2$, a large fraction ($\approx 0.285$) of pairs of states were in fact C-optimal. The sampling of states is such that each of the real states is chosen from a uniform distribution on the unit 3-sphere $S^3$, which simply arises from normalization of the four real components.  Strong evidence was provided that the number $0.285 \ldots $, obtained initially from numerical simulations, was in fact $(\pi-2)/4$. Below it is shown that this is in fact $p^{(2)}_{2,2}-1/2$, whose evaluation then confirms the result.
For completeness we state that when $k=3$ about $5.12\%$ of triples were C-optimal while it was also shown that there was not even {\it one} quadrapulet of real states that were so. Therefore the set of complex states is necessary for there to be C-optimal states in general. For $k>2$ though, there does not seem to be a direct connection to the problem of products of random matrices.

If $|\phi_1\kt$ and $|\phi_2\kt$ are an optimal pair satisfying Eq.~(\ref{coptimal}) we refer to them below as ``co-optimal". Such optimal pairs satisfy the following conditions \cite{Shuddhodhan2011}: 
\begin{multline}
r_{11}r_{22}>0, \;\; \mbox{and}\;\; r_{11}r_{22}-r_{12}^2<0, \\ \;\; \mbox{where}  \;\; r_{ij}=\br \phi_i |\sigma_y \otimes \sigma_y |\phi_j \kt.
\end{multline}
Here $\sigma_y$ is one of the Pauli matrices. If $|\phi_i\kt$ is a real state of two qubits, the concurrence $C\left(|\phi_i \kt \br \phi_i|\right)=|r_{ii}|$.
Let $|\phi_1\kt =(|00\kt +|11\kt )/\sqrt{2}$, be an maximally entangled state, so that $C \left(|\phi_1 \kt \br \phi_1|\right)=1$. 
What characterizes states that are co-optimal with this? If 
\beq |\phi_2\kt = a |00\kt +b |01\kt +c |10\kt +d | 11\kt,
\eeq
be such a state, the conditions stated above lead to $r_{11}=-1$, 
$r_{22}=2(bc-ad)$ and $r_{12}=(a+d)/\sqrt{2}$. This implies the following:
\beq
ad-bc >0, \;\; (a+d)^2 - 4 (ad-bc) >0,
\eeq 
which beg to be formulated as conditions on the matrix of coefficients 
\beq
M_1= \left( \begin{array}{cc} a &b \\c&d \end{array} \right),
\eeq 
being equivalent to the requirement that $\det(M_1)>0$ and $M_1$ has only {\it real} eigenvalues. Of course if $\det(M_1)<0$, the eigenvalues are anyway real. Thus a state is co-optimal with a maximally entangled state if its matrix of co-efficients has a positive determinant, yet has real eigenvalues.


Formulated as above, the fraction of states that are co-optimal with a maximally entangled state,  is closely allied to the question of the fraction of $2 \times 2 $ real matrices that have real eigenvalues. The matrix elements can be drawn from a normal i.i.d. random process, such as $N(0,1)$. That this gives us the same answer as sampling uniformly from the normalization sphere of $|\phi_2 \kt$ is evident, as the question of reality of eigenvalues of a matrix remains independent of overall multiplication by scalars. Thus we get the fraction $f_{\pi/4}$ of states that are co-optimal with a maximally entangled state to be
\beq
f_{\pi/4}= p_{2,2}-\f{1}{2}=\df{1}{\sqrt{2}} -\df{1}{2}.
\eeq
The $-1/2$ arises as the fraction $p_{2,2}$ will also include {\it all} instances when $\det(M_1)<0$, which are to be subtracted, and $\det(M_1)$ is equally likely to be positive or negative.  Thus about $20.7\%$ of real states are co-optimal with a maximally entangled one.


To generalize the above, consider one state as $|\phi_1\kt = \cos \ta |00 \kt +  \sin \ta |11 \kt$, $0 \le \ta \le \pi/4$, and the measure $f_{\ta}$ of states that are co-optimal with it, $\ta=\pi/4$ being what was just discussed. That any one state of the pair can be chosen such as this follows from Schmidt decomposition. The uniform (Haar) distribution  on the normalization sphere $S^3$ induces an invariant measure, say $\mu (\ta)$. Then the fraction of pairs of states that are co-optimal is given by
\beq
\label{avrg}
\br f\kt=\int_0^{\pi/4} f_{\ta} \mu(\ta) d\ta.
\eeq

The conditions of co-optimality of $|\phi_1\kt$  and a general real two-qubit state $|\phi_2\kt$, with $r_{11}=-\sin 2 \ta$, $r_{22}=2 (bc-ad)$ and $r_{12}=a \cos \ta +b \sin \ta$, now translate to those on the product:
\beq
\label{prod2}
M_2= \left( \begin{array}{cc} \cos \ta &0 \\0&\sin \ta \end{array} \right) \left( \begin{array}{cc} a &b \\c&d \end{array} \right),
\eeq 
as $\det(M_2)>0$ and $M_2$ has real eigenvalues. Once again, in this problem it is equivalent to assume $(a,b,c,d)$ are i.i.d. $N(0,1)$ numbers or uniformly distributed on the sphere $a^2+b^2 +c^2+d^2=1$.

Quite independent of the discussion above, but equivalently, one may start with a product of two random matrices, say $A_1 A_2$, and perform a singular value decomposition of $A_1$ to get the product $O_1 \Lambda_1 O_2^T A_2$=$O_1 \Lambda_1 O_2^T A_2O_1 O_1^T$=
$O_1 \Lambda_1 {\tilde A_2} O_1^T$. Evidently the spectrum of the original product is same as that of $\Lambda_1 {\tilde A_2}$. Here $O_1$ and $O_2$ are orthogonal matrices and $\Lambda_1 $ is a diagonal matrix with positive elements, and ${\tilde A_2}=O_2^T A_2O_1$.
Observe that if the elements of a matrix $A$ are i.i.d. $N(0,1)$ distributed, those of the products
$OA$, and $AO$, where $O$ is an arbitrary orthogonal matrix, are also identically distributed. Therefore it follows that ${\tilde A_2}$ has elements that are  i.i.d. $N(0,1)$ distributed. Hence one may well begin with the product in Eq.~(\ref{prod2}) without any loss of generality. 

That the diagonal elements can be so taken, with $0 \le \theta \le \pi/4$, and distributed naturally according to the measure $\mu(\theta)= 2 \cos 2 \ta$ is now shown. The eigenvalues of $A_1 A_1^T$, can be chosen as $\lambda > 0$  and $1-\lambda \le \lambda$, as division by an overall number, the $\tr(A_1A_1^T)$, does not affect the nature of the reality of the eigenvalues of the product $A_1A_2$. With $A_1$ having elements drawn from an i.i.d. normal random process, namely the Ginibre ensemble, but with the trace restricted to unity, the distribution of the eigenvalues 
of $A_1 A_1^T$ is known in general \cite{Sommers2001,BengtssonZyczkowski}. For the special case of $2$ dimensional matrices, the distribution of the 
larger value can be read off as $P(\lambda)= (2 \lambda-1)/\sqrt{\lambda(1-\lambda)}$. Hence with the parametrization that $\lambda= \cos^2 \ta$
as above, the singular value being $\cos \ta$, the distribution $\mu(\ta)$ follows immediately.

Let $p_{\ta}$ be the fraction of matrices $M_2$ that have real eigenvalues as $(a,b,c,d)$ are taken from $N(0,1)$ and $\ta$ is fixed. This is realized each time the discriminant $\Delta_2=(a \cos \ta +d \sin \ta)^2 -4 \sin \ta \cos \ta (ad-bc) \ge 0$. This is rewritten as  $\Delta_2=(a \cos \ta -d \sin \ta)^2 + 2\sin 2\ta \,bc \ge 0$, which is a condition on the sum of two statistically independent quantities. Using the fact that $x=(a \cos \ta -d \sin \ta)$ is distributed according to $N(0,1)$ for all $\ta$ enables the following form:
\beq
p_{\ta}= \int_{-\infty}^{\infty} \Theta\left( \f{\beta}{2} x^2+ y z \right) e^{-(x^2+y^2+z^2)/2} \df{dx dy dz}{(2 \pi)^{3/2}},
\eeq
where $\beta=1/\sin 2 \ta$. Note that as $\ta \rarrow 0$, $\beta \rarrow \infty$, and $p_{\ta } \rarrow 1$. Taking the derivative with respect to $\beta$ converts the Heaviside step function 
into a Dirac delta function. Effecting a series of simplifications thereafter, including using polar coordinates for $y$ and $z$ results in the following remarkably simple equation:
\beq
\df{d p_{\ta}}{d \beta} = \f{1}{4 \pi} \int_0^{\pi} \df{\sqrt{\sin \phi } \,d \phi}{(\sin \phi +\beta)^{3/2}}.
\eeq

Integrating with respect to $\beta$ and incorporating the boundary condition at $\ta=0$ gives:
\beq
\label{intandsum}
\begin{split}
&p_{\ta}=1-\f{1}{2 \pi} \int_0^{\pi} \sqrt{\df{\sin \phi  }{\sin \phi +\beta}} \,d \phi\\
&=1-\f{1}{2 \pi} \sum_{k=0}^{\infty} \f{(-1)^k}{k!}\df{\Gamma(k+\f{1}{2}) \Gamma(\f{k}{2}+\f{3}{4})}{ \Gamma(\f{k}{2}+\f{5}{4})} (\sin 2 \ta)^{k+\f{1}{2}}.
\end{split}
\eeq
The integral in this equation does not seem to acquire a simple form except when $\beta=1$, which corresponds to $\ta=\pi/4$ and gives $p_{\pi/4}=1/\sqrt{2}$, in  agreement with the known result, stated previously as $p_{2,2}$. It follows that $f_{\ta}=p_{\ta}-\tfrac{1}{2}$ is the fraction of states that are co-optimal with the state $\cos \ta |00\kt +\sin \ta |11\kt$. Note that as the concurrence in this states is $\sin 2 \theta$, the fraction $f_{\ta}$ is simply a function of this. One can now use Eq.~(\ref{avrg}) to find the fraction of co-optimal pairs. The required invariant measure being $\mu(\ta )=2 \cos 2\ta$, is most well suited to express $\br f \kt $ as an infinite series as in Eq.~(\ref{intandsum}), that may be identified with generalized hypergeometric functions.

\begin{figure}
\begin{center}
\includegraphics[scale=1.3]{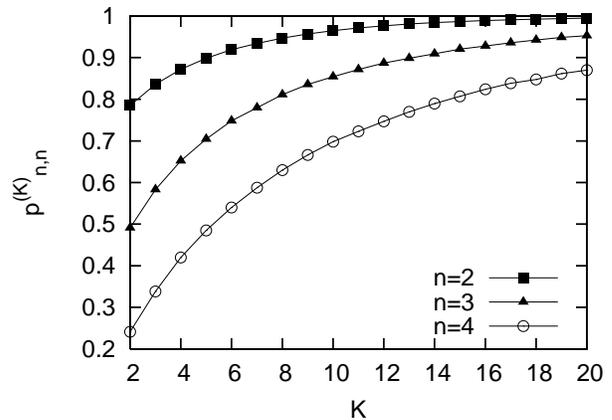}
\caption{The probability that all eigenvalues of a product of $K$ random $n$ dimensional matrices are real, based on 100,000 realizations. } 
\label{pk22} 
\end{center}
\end{figure} 

Equivalently one may use the integral in this equation to express $\int_0^{\pi/4} p_{\ta} \mu(\ta) d\ta$, the probablity that the product of two random $2\times 2$ matrices has real eigenvalues as
\beq
\label{theintegral}
p^{(2)}_{2,2}=\f{1}{2 \pi} \int_0^{\pi} \df{\sinh^{-1}(\sqrt{\sin \phi})}{\sin \phi} \, d \phi.
\eeq
this follows as the $\ta$ integral can be carried out in an elementary way, and also from the 
evaluation $\int_0^{\pi} \sqrt{(1+\sin x)/\sin x} \,dx =2 \pi$. The integral in Eq~(\ref{theintegral})  does not appear to be in standard tables, nor fully evaluated by mathematical packages, but as indicated from previous work it is in fact simply $\pi^2/2$. Therefore it seems interesting enough to warrant a more complete evaluation. The expansion of the inverse hyperbolic functions enables the integral to be written as:
\begin{multline}
\f{1}{\sqrt{2 \pi}} \Gamma^2\left(\f{1}{4}\right) \pFq{3}{2}{\f{1}{4},\f{1}{4},\f{1}{4} }{\f{1}{2},\f{5}{4}}{1}-\\\f{1}{24 \sqrt{2 \pi}} \Gamma^2\left(-\f{1}{4}\right) \pFq{3}{2}{\f{3}{4},\f{3}{4},\f{3}{4} }{\f{3}{2},\f{7}{4}}{1}.
\end{multline}
Both the generalized hypergeometric functions appearing here are of the Saalsch\"utz type, the sum of the top rows being 1 less than the sum of the bottom. Theorem 2.4.4 in \cite{AndrewsAskeyRoy} can be evoked for such functions, and it is remarkable that this is precisely the form of the R.H.S. of the identity therein, which results in its evaluation as 
\beq
\sqrt{2}\, \pi \; \pFq{2}{1}{\f{1}{4},\f{1}{4}}{\f{5}{4}}{1} =\f{\pi^2}{2}.
\eeq
Here use is made of an identity of Gauss for ${}_2F_1$ at arguments of unity \cite{AndrewsAskeyRoy}, and leads to $p^{(2)}_{2,2}=\pi/4$, and hence finally $\br f\kt =(\pi-2)/4$.

\begin{figure}
\begin{center}
\includegraphics[scale=1]{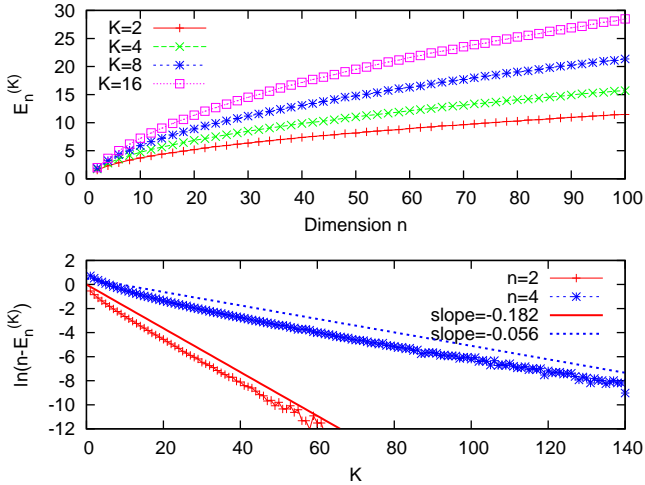}
\caption{The expected number of real eigenvalues of products of $K$ random $n$ dimensional matrices, based on 100,000 realizations. The top panel shows variation with the $n$ for fixed values of $K$, while the bottom one displays the exponential approach of the expected number of real eigenvalues to the dimension of the matrix as $K$ increases.} 
\label{expected} 
\end{center}
 \end{figure}

\begin{figure}
\begin{center}
\includegraphics[scale=1.3]{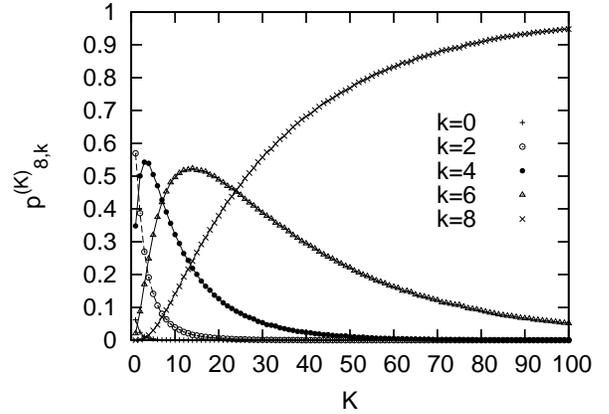}
\caption{The probability that $k$ eigenvalues of a product of $K$ random $8$ dimensional matrices are real, based on 100,000 realizations. The $k=0$ case is barely seen in this scale.} 
\label{pk8k} 
\end{center}
\end{figure}

The generalizations, dealt with numerically below, are to products of more that two $2\times 2$ matrices as well as to higher dimensional matrices and for a variable number of products. The behavior of $p^{(K)}_{2,2}$ for $K\ge2$ is seen in Fig~(\ref{pk22}) and shows this monotonically increasing with $K$. In the same figure is also shown the corresponding probabilities that all the eigenvalues are real for such products of 3 and 4 dimensional matrices. This increase in the probability that all eigenvalues are real is also reflected in the expected number of real eigenvalues. This is shown in Fig.~(\ref{expected}) where this number: $E_n^{(K)}=\sum_{k=0}^n k p^{(K)}_{n,k}$ is plotted as a function of $n$ for fixed values of number of products $K$ in the top panel. In the bottom panel the expected number is shown as a function of $K$ for 2 and 4 dimensional matrices. An exponentially fast approach to the dimension of the matrix is observed here. Thus $n-E_n^{(K)} \sim \exp( -\gamma_n K)$ seems to hold with $\gamma_n$ decreasing with increasing $n$ ($\gamma_2$ and $\gamma_4$ are the slope values in Fig.~(\ref{expected})).  Numerical results also indicate that the probability that there are $k$ real eigenvalues for $k<n$, while not necessarily monotonic, does eventually vanish with the number of products, leaving the dominant case as the one with {\it all} eigenvalues real. This is illustrated in Fig.~(\ref{pk8k}) where the quantity $p^{(K)}_{8,k}$ is plotted for $8$ dimensional matrices for $k=0,2,4,6,$ and $k=8$, corresponding to the probability of finding $k$ real eigenvalues. A similar trend is observed for all other numerically tested dimensionalities, and hence there is a strong case that this is true in general.

\begin{figure}
\begin{center}
\includegraphics[scale=.7]{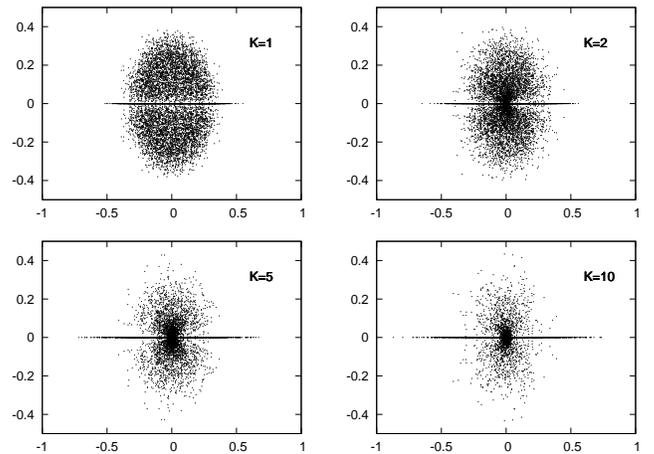}
\caption{The eigenvalues of $K$ products of 10 dimensional random matrices, after they have been divided by the corresponding Frobenius norms. The real and imaginary parts are plotted for 1000 realizations of such products.} 
\label{evals} 
\end{center}
\end{figure}

For a fixed dimensionality as the number of products increases more eigenvalues ``condense" from the complex plane onto the real axis. The distribution of the eigenvalues hence changes significantly as well. For a single random matrix the eigenvalues are asymptotically distributed according to the circular law \cite{Girko1984,Bai1997}, while the real eigenvalues are asymptotically uniformly distributed \cite{Edelman1997}. In Fig.~(\ref{evals}) are shown the eigenvalues of products of $10$ dimensional real matrices. This is shown for $4$ values of $K$, namely 1, 2, 5, and 10, and the distortion from an approximately circular law is evident with the formation of two lobes. The eigenvalues are divided by the norm of the products of the matrices so that the values are not exponentially increasing and can be compared.

In conclusion and summary,  the number of real eigenvalues for products of real random matrices has been studied.  The case of products of two 2 dimensional random matrices was fully analytically solved and it was shown that in the fraction of $\pi/4$ cases, the matrices had real eigenvalues.  This solved a problem of entanglement, where it was shown that the fraction of optimal pairs of two qubit states is therefore $(\pi-4)/2$. Generalizations show that with increasing number of products, {\it all} the eigenvalues tend to be real with probability approaching unity. This moreover seems valid for all matrix dimensions. Needless to say, the numerical results pose interesting challenges, as the resulting matrices have highly correlated matrix elements. For one thing, {\it why } all the eigenvalues become real is a natural question, and while admittedly tentative, it might have to do with the rapid convergence of vectors under repeated multiplication by random matrices. Thus it is also interesting to explore connections, if any, between the exponential rates found in this paper and Lyapunov exponents.

\acknowledgements It is a pleasure to than Karol Zyczkowski for long standing discussions about the optimality problem and related issues.

\bibliography{QIpRMT}

\end{document}